\documentclass[twocolumn,prd,aps,showpacs,showkeys,amsmath,amssymb,nofootinbib]{revtex4-1}
\usepackage{bm}

\usepackage{color}
\usepackage{amsmath}
\usepackage{amsfonts}
\usepackage{verbatim}
\usepackage{amssymb}
\usepackage{graphicx}
\usepackage{epstopdf}
\usepackage{mathrsfs}
\usepackage{epsfig}
\usepackage{slashed}
\usepackage{bbold}
\usepackage{color} 

\begin{document}


\title{Spin 2 Quasinormal Modes in Generalized Nariai Spacetimes}
\author{Jo\'as Ven\^ancio and Carlos Batista}
\email[]{carlosbatistas@df.ufpe.br}
\affiliation{Departamento de F\'{\i}sica, Universidade Federal de Pernambuco,
Recife, Pernambuco  50740-560, Brazil}


\begin{abstract}
In this work we analytically obtain the quasinormal spectrum for the gravitational perturbation on a higher-dimensional generalization of the Nariai spacetime that is comprised of the direct product of the two-dimensional de Sitter space with several two-spheres. A key step in order to attain this result is to use a suitable basis for the angular functions depending on the rank of the tensorial degree of freedom that one needs to describe. Here we define such a basis, which is a generalization of the tensor spherical harmonics that is suited for spaces that are the product of several spaces of constant curvature.
\end{abstract}

\maketitle


\section{Introduction}


Suppose one disturbs a system that is initially at stable equilibrium. The forces driven by such disturbance are generally nonlinear on the perturbation. However, assuming small perturbations, it turns out that in several problems the components of the forces that are linear on the disturbance amplitude are much more relevant than the higher order terms, so that the dynamical equation can be linearized. This is tantamount to approximating a general potential by a parabola that osculates its minimum. In this scenario, the so-called normal frequencies are of central importance. These are the natural oscillation frequencies of the system, irrespective of the details of the disturbance. When the perturbations naturally tend to decay as time passes by, these frequencies are complex numbers whose real parts give the oscillation frequencies while the imaginary parts are related to the characteristic times of the decay. These complex frequencies form the so-called quasinormal spectrum of the system.

The study of perturbations is of central importance in almost all branches of physics, since often the physical systems are in a stable configuration and the changes are all due to small disturbances that do not build up as time passes by. The perturbation formalism is even more necessary when the mathematical equations that describe the dynamics of a system are nonlinear, since the effect of perturbations can generally be handled by means of linear equations, providing thus a great deal of simplification. Relativistic gravitational physics is an important example of this, since its field equation, Einstein's equation, is a coupled set of ten nonlinear partial differential equations that are impossible solve analytically in the generic case, i.e. without assuming the existence of special symmetries. However, one can start with a stable solution, like Schwarzschild or Kerr black holes, and then study the dynamical evolution of arbitrary perturbations in these backgrounds. Since in the classical realm the event horizon is a one way membrane, dissipation is always present in black hole scenario, so that the natural frequencies of these gravitational systems are complex \cite{Vishveshwara70a}. Due to the recent capability of detecting gravitational waves \cite{Abbott}, the study of quasinormal modes of the gravitational field got even more attention lately and, certainly, will increase its relevance in the forthcoming years. For instance, by measuring the natural frequencies of a black hole through a gravitational wave detector one can find the parameters of this black hole, like mass and angular momentum. Moreover, one can test alternative theories of gravity against the experiment \cite{Konoplya:2016pmh}.  For very good reviews on quasinormal modes in gravitational physics, the reader is referred to \cite{Berti09,Konoplya:2011qq,Kokkotas,Nollert99,Cardosotese}.

In spite of the fact that the perturbation equation for the gravitational field is much more simple to solve than the full Einstein's equation, even the simplest cases, like perturbations on the Schwarzschild background, proved to be challenging. In fact, first and foremost, the quasinormal spectrum of Schwarzschild black hole cannot be obtained exactly \cite{NollertNnumeric,Natario:2004jd}, analytical results were found only for the extremal case of Schwarzschild-de Sitter solution. Moreover, this humble problem  was a source of some debate and confusion in the literature. Regge and Wheeler were the first to decompose the gravitational perturbations in Schwarzschild background in terms of tensor harmonics \cite{Regge}, a tensorial generalization of the spherical harmonics. In order to perform this, they classified gravitational perturbations into two types: odd parity and even parity. In addition, they showed that the equations for the odd parity perturbations can be put into the form of a Schr\"odinger-like equation with a nonintegrable potential. However, there were some minor errors in the equations given by Regge and Wheeler. Indeed, Manasse pointed  out  that  the  equations  appearing  in the  literature  contained  mistakes  and  were  inconsistent  with Einstein's  field equation \cite{Manasse}. Brill  and  Hartle rederived  the  odd  parity  equations  which  once again  contained  some  errors  as  published \cite{Brill}. The correct differential equations for perturbations on the Schwarzschild metric, for both parities, have been given by Vishveshwara \cite{Vishveshwara67}, displayed in Appendix A of his doctoral thesis and published latter in Ref. \cite{Vishveshwara70b}. Using the latter equations, Zerilli \cite{Zerilli} has found that the even parity perturbation equations can also be put into a Schr\"odinger-like equation, just as the odd parity perturbations. Zerilli's equation yields an enormous simplification in the analysis of such perturbations and his work was of great significance in the study of gravitational radiation formed from an asymmetric gravitational collapse. It is also worth mentioning the contribution of Fackerell on the analysis of the solutions to Zerilli's equation \cite{Fackerell}. The Regge-Wheeler formalism was later extended to other static black holes in four dimensions \cite{Zerilli, Moncrief01, Moncrief02} and in higher dimensions \cite{Takahashi, Kodama}. Rotating black holes in four dimensions were tackled in the seminal works of Teukolsky \cite{Teukolsky72, Teukolsky74}. Recently, some techniques based on monodromy calculations have been put forward to obtain analytical expressions for the quasinormal spectrum of perturbations in five-dimensional Kerr background \cite{Barragan1}. However, the latter spectrum is written in terms of transcendental equations whose solutions must be found numerically \cite{Barragan-Amado:2018pxh}.

Continuing the study of the gravitational perturbation in higher-dimensional spacetimes, here we look for the quasinormal spectrum of a generalization of Nariai solution. In fact, we manage to obtain analytically the spectrum for the background discussed in Ref. \cite{Carlos1}, see also Ref. \cite{VitorN}. In four dimensions the spacetime considered here reduce to the well-known Nariai solution, which is a static solution of Einstein's field equation that can be attained from the Schwarzschild-de Sitter black hole in the limit in which the cosmological and black hole horizons coincide \cite{Nariai1,Nariai2}. Thus, our spectrum should reduce to Nariai quasinormal frequencies when the dimension is set to four. This turn out to be true when a comparison with the spectrum displayed in Ref. \cite{Cardoso03} is performed, although the explicit calculation is not shown there. Nevertheless, our result in four dimension coincides only partially with the ones obtained in Refs. \cite{Vanzo, Ortega09}, where it is shown three types of spectra for the gravitational perturbation in Nariai spacetime depending on the tensor nature of the degree of freedom being perturbed, namely scalar, vectorial or tensorial. On the other hand according to our calculations all these degrees of freedom must have the same spectrum in four dimensions. More precisely, our spectrum coincide with two of the three spectra considered in Refs. \cite{Vanzo, Ortega09}. We believe that a typo have occurred in the first of these articles, Ref. \cite{Vanzo}, and then have been propagated in Ref. \cite{Ortega09}.

The outline of this article is the following. In Sec. \ref{Sec.Problem} we present the problem and its details. First we discuss the general field equation for the gravitational perturbation and then present the higher-dimensional background that is adopted here. Moreover, we discuss the symmetries of such background and illustrate how we can take advantage of these to find a suitable basis to expand the components of the gravitational perturbation. In particular, we split the degrees of freedom into two broad types that are not mixed by the field equations, namely the odd perturbations and the even perturbations. Then, in Sec. \ref{Sec.Odd} we consider the equations obeyed by the odd degrees of freedom. After imposing suitable boundary conditions, we are then able to find an analytical expression for the frequencies that are compatible with such boundary conditions. In Sec. \ref{Sec.Even} we follow analogous steps for the even degrees of freedom. Finally, in Sec. \ref{Sec.Conclusion} we present some conclusions and, taking into account the results of a previous work of ours \cite{JoasI}, we find a formula for the quasinormal spectrum of a perturbation with generic spin and mass in the higher-dimensional Nariai space considered here. In the brief Appendix \ref{Sec.Appendix}, we provide explicit expressions for the angular functions adopted as a basis in this work.


\section{Field Equation for the Spin 2 Perturbation}\label{Sec.Problem}

Let the metric $g_{\mu\nu}$ be a solution to Einstein's vacuum equation with a cosmological constant $\Lambda$, namely
\begin{equation*}
  \mathcal{R}_{\mu\nu} = \Lambda \,g_{\mu\nu}   \,,
\end{equation*}
where $\mathcal{R}_{\mu\nu}$ is the Ricci tensor associated to the metric. Then, performing a perturbation on the gravitational field $\tilde{g}_{\mu\nu}$ and on some matter fields $\tilde{\Phi}_{i}$ living in this perturbed background, so that
\begin{equation}\label{PerturbPhi}
\tilde{g}_{\mu\nu}\,=\,g_{\mu\nu}\,+\,h_{\mu\nu}\; , \;\textrm{ and } \quad  \tilde{\Phi}_{i} \,=\,0 + \phi_{i} = \phi_{i}\,,
\end{equation}
where $h_{\mu\nu}$ and $\phi_i$ are infinitesimal, it follows that the matter field perturbation decouples from the gravitational perturbation. Indeed, the matter fields appear in Einstein's equation through its energy-momentum tensor, which is typically quadratic or of higher order in the matter fields $\tilde{\Phi}_{i}$. Since we are assuming that these were zero, before the perturbation,  it turns out that quadratic expressions on $\tilde{\Phi}_{i}$ are quadratic on the infinitesimal perturbation and, therefore, should be neglected. Thus, the differential equation for $h_{\mu\nu}$ does not have the matter fields as a source. Likewise, in the field equations for $\tilde{\Phi}_{i}$  we should only consider the non-perturbed metric $g_{\mu\nu}$, since each term of these equations are at least linear in the infinitesimal field $\phi_i$, so that any contribution of $h_{\mu\nu}$ would lead to a negligible order. Thus, in general, the perturbations of mater fields can be analysed independently from the gravitational perturbation. In our previous work \cite{JoasI} we have considered quasinormal modes associated the matter fields of spin $0$, $1/2$, and $1$ in the generalized Nariai spacetime.  Now, in this paper, we will tackle perturbations on the  spin $2$ gravitational field.

Linearizing Einstein's equation around the background with metric $g_{\mu\nu}$ we end up with the following equation for $h_{\mu\nu}$:
\begin{align}
  2\nabla^{\sigma}\nabla_{(\mu}h_{\nu)\sigma}\,-\,\Box h_{\mu\nu} \,-\, \nabla_{\mu}\nabla_{\nu}\,h - \,2\Lambda \ h_{\mu\nu} = 0\,, \label{Eqh}
\end{align}
where $\nabla_\mu$ is the Levi-Civita covariant derivative built from the initial background metric $g_{\mu\nu}$, $h = h^{\mu}_{\phantom{\mu}\mu}$, and $\Box = \nabla^\mu\nabla_\mu$, with the indices being raised using the inverse of the unperturbed metric, $g^{\mu\nu}$.

Here, the background spacetime is assumed to be a higher-dimensional generalization of the Nariai spacetime that is the direct product of the de Sitter space $dS_{2}$ with $(d-1)$ spheres $S^{2}$. The line-element of this $2d$-dimensional spacetime is given by
\begin{equation}\label{nariai-metric}
ds^2 = -(1-\Lambda r^{2})\,dt^{2} + \frac{dr^{2}}{1-\Lambda r^{2}}+ \frac{1}{\Lambda}\sum_{j=2}^{d}\,d\Omega_{j}^{2} \,,
\end{equation}
where $d\Omega_{j}^{2}$ is the line element of the $j$-th unit sphere, given explicitly by
\begin{equation}\label{functionf1}
 d\Omega_{j}^{2}\,=\,d\theta_{j}^{2}\,+\, \text{sin}^{2}\theta_{j}\,d\phi_{j}^{2}   \,.
\end{equation}
For each of the $(d-1)$ spheres, there exist three independent Killing vectors that generate rotations, namely
\begin{equation*}
  \left\{
     \begin{array}{ll}
        \mathbf{k}_{1,j} = \sin\phi_j\, \partial_{\theta_j} + \cot\theta_j \cos\phi_j\,\partial_{\phi_j}\,,\\
       \mathbf{k}_{2,j} = \cos\phi_j\, \partial_{\theta_j} - \cot\theta_j \sin\phi_j\,\partial_{\phi_j} \,,\\
       \mathbf{k}_{3,j} = \partial_{\phi_j}  \,.
     \end{array}
   \right.
\end{equation*}
In particular, note that the operator that acts on $h_{\mu\nu}$ in Eq. (\ref{Eqh}) commutes with the Lie derivatives $\mathcal{L}_{\mathbf{k}_{I,j}}$. Indeed, since $\mathbf{k}_{I,j}$ are Killing vector fields of the background metric, it follows that the action of $\mathcal{L}_{\mathbf{k}_{I,j}}$ on $g_{\mu\nu}$ yields zero. Since the Levi-Civita covariant derivative depends only on the background metric it follows that
$\mathcal{L}_{\mathbf{k}_{I,j}}\nabla_\mu = \nabla_\mu \mathcal{L}_{\mathbf{k}_{I,j}}$. Thus, since $\mathcal{L}_{\mathbf{k}_{I,j}}$ generates infinitesimal rotations in the $j$-th sphere, it turns out that if $h_{\mu\nu}$ is a solution of Eq. (\ref{Eqh}), then its rotated version will also be a solution. This humble assertion has an important practical consequence, namely when we expand $h_{\mu\nu}$ in terms of irreducible representations of the $SO(3)$ symmetry group associated to each sphere, we just need to consider the elements of the representation basis with $m_j=0$, where $m_j$ is the eigenvalue with respect to $\mathbf{k}_{3,j}$. The other possible values for $m_j$ can be attained by applying the ladder operators, which are just linear combinations of the rotations generated by $\mathbf{k}_{1,j}$ and $\mathbf{k}_{2,j}$. This leads to great simplification in the calculations. We shall return to this point when we introduce the basis used to expand the components of $h_{\mu\nu}$.

In addition to these Killing vectors, $\mathbf{k}_t = \partial_t$ also generates an isometry. In particular, this Killing vector is light-like at the closed surfaces $r = \pm \Lambda^{-1/2}$, so that these are Killing horizons. The boundary conditions of the quasinormal modes will be posed at these surfaces, as discussed in Ref. \cite{JoasI}, and the domain of interest will be $r\in ( -\Lambda^{-1/2}, \Lambda^{-1/2})$. In such domain one can use the tortoise coordinate  $x$ defined by $r = \Lambda^{-1/2} \tanh(x\Lambda^{1/2})$, in terms of which the line element becomes
\begin{equation}\label{nariai-metric2}
g_{\mu\nu}dx^{\mu}dx^{\nu} = f(x)\left( -\,dt^{2} + dx^{2}\right) + \frac{1}{\Lambda}\sum_{j=2}^{d}\,d\Omega_{j}^{2} \,,
\end{equation}
where $f = f(x)$ is the following function of the coordinate $x$:
\begin{equation*}
 f = \textrm{sech}^2(x\sqrt{\Lambda}) \,.
\end{equation*}

Besides the continuous symmetries generated by Killing vectors, there are also some discrete symmetries. In particular, the line element is invariant under the parity transformation (spatial inversion) in each of the spheres. More precisely, the changes
\begin{equation}\label{parityT}
  \theta_j \,\rightarrow\, \pi - \theta_j \;\;\;  \textrm{and} \;\;\; \phi_j \,\rightarrow\,\phi_j + \pi
\end{equation}
do not modify the line element (\ref{nariai-metric}). Denoting this transformation by $P_j$, it follows that $P_j^2$ is the identity transformation, so that the eigenvalues of this transformation are $\pm1$. Objects unchanged under $P_j$ (eigenvalue $+$1) are said to have even parity, while those that change by a global sign (eigenvalue $-$1) are said to have odd parity. It turns out that the differential operator that acts on $h_{\mu\nu}$ in Eq. (\ref{Eqh}) commutes with these parity transformations, so that the components of $h_{\mu\nu}$ with different parities will not mix in this equation. Thus, in order to integrate Eq. (\ref{Eqh}), we can analyse the even and odd parts of $h_{\mu\nu}$ separately without loosing generality. This fact will be of great practical relevance in what follows.

Just for sake of illustration, suppose that we would like to study a test massless scalar field $\Phi$ propagating in the 4-dimensional Nariai spacetime, so that in our notation $d=2$. Then, in order to integrate its equation $\Box \Phi = 0$,
 it is quite useful to assume that this field has the form
\begin{equation}\label{scalar}
  \Phi(t,x,\theta_2,\phi_2) = e^{i\omega t} \,\phi(x) \,Y_{\ell_2}^{m_2}(\theta_2,\phi_2) \,.
\end{equation}
The latter time dependence is due to the fact that $\partial_t$ is a Killing vector, so that $t$ appears in the equation $\Box \Phi = 0$ just through derivative operators $\partial_t$. In turn, $Y_{\ell_2}^{m_2}$ is a spherical harmonic that is suitable to be used as a basis for the angular dependence due to the fact that the background has spherical symmetry. The index $\ell_2$ labels the irreducible representations of the $SO(3)$ isometry subgroup associated to the spherical part of the line element, while $m_2$ is an integer in the domain $-\ell_2\leq m_2\leq\ell_2$ that labels the $(2\ell_2+1)$ elements of the basis of the irreducible representation $\ell_2$. It is worth stressing that the most general solution for the scalar field is not the one given in Eq. (\ref{scalar}), but rather a general linear combination of the solution (\ref{scalar}) for different frequencies $\omega$ and different values of the separation constants $\ell_2$ and $m_2$.

Now, suppose that we are interested in integrating the field equation for a spin 1 gauge field
$A_\mu$ in this 4-dimensional background, $\nabla^\mu \nabla_{[\mu}A_{\nu]}=0$. Since the spherical harmonics are a basis for the functions in the sphere, we could also expand $A_\mu$ in terms of them just as we did in Eq. (\ref{scalar}), namely
\begin{equation}\label{A1}
  A_\mu(t,x,\theta_2,\phi_2) = e^{i\omega t} \,\Delta_\mu(x) \,Y_{\ell_2}^{m_2}(\theta_2,\phi_2) \,.
\end{equation}
However, this is not the most suitable choice. Indeed, while the components $A_t$ and $A_x$ are scalars with respect to the action of the isometry subgroup $SO(3)$, the components $A_{\theta_2}$ and $A_{\phi_2}$ transform under rotations as the components of a covector field in the sphere. Therefore, the most natural  way to expand $A_{\theta_2}$ and $A_{\phi_2}$ is using a basis of 1-forms in the sphere. Starting from the spherical harmonic $Y_{\ell_2}^{m_2}(\theta_2,\phi_2)$, which are scalar fields in the sphere, one can take covariant derivatives and build the following 1-forms
\begin{equation}\label{V2}
  V^{+}_{a_2} = \hat{\nabla}_{a_2}Y_{\ell_2}^{m_2} \; , \;\; \textrm{and} \;\; V^{-}_{a_2}  = \hat{\epsilon}_{a_2c_2}\hat{\nabla}^{c_2}Y_{\ell_2}^{m_2} \,,
\end{equation}
where the indices $a,b,c$ run through $\{\theta,\phi\}$, $\hat{\nabla}_{a_2}$ denotes the covariant derivative in the unit sphere, whose line element and metric $\hat{g}_{a_2b_2}$ are given by
\begin{equation*}
  ds^2 = \hat{g}_{a_2b_2} dx^{a_2}dx^{b_2} = d\theta_2^2 +  \sin^2\theta_2\,d\phi_2^2\,,
\end{equation*}
whereas $\hat{\epsilon}_{a_2b_2}$ is the volume form in the sphere. Explicit expressions for $V^{\pm}_a$ are provided in appendix \ref{Sec.Appendix}.
These two 1-forms have different behaviours under the parity transformation (\ref{parityT}). Since a spherical harmonic transforms as
\begin{equation}\label{SphericalHarmParity}
 Y_{\ell_2}^{m_2} \xrightarrow{\textrm{parity}} (-1)^{\ell_2}\,Y_{\ell_2}^{m_2}
\end{equation}
under the parity transformation (\ref{parityT}), it follows that the 1-form $V^{+}=V^{+}_{a_2}dx^{a_2}$ is multiplied by $(-1)^{\ell_2}$ as well, while $V^{-}=V^{-}_{a_2}dx^{a_2}$ gets multiplied by $(-1)^{\ell_2+1}$. In what follows, given a certain irreducible representation $\ell$, objects that transform under parity in the same way as the spherical harmonics, namely as in Eq. (\ref{SphericalHarmParity}), are dubbed even, while objects that get an extra minus sign under parity transformation are called odd.
Thus, for instance, we shall say that $V^{+}$ has even parity, while $V^{-}$ has odd parity. Using these objects, the natural decomposition for $A_\mu$ is
\begin{align*}
   A_\mu dx^\mu = e^{i\omega t} & \big[  \left( \Delta_t \,dt+ \Delta_x \ dx\right) Y_{\ell_2}^{m_2}  \\
&\;\;\;\;\;\; + \Delta_+ \,V^{+}  + \Delta_- \, V^{-}   \big] \,,
\end{align*}
where $\Delta_t$, $\Delta_x$, and $\Delta_{\pm}$ are functions of the radial coordinate $x$. These four functions accounts for the four degrees of freedom of a 1-form in four dimensions. Moreover, in the field equation $\nabla^\mu \nabla_{[\mu}A_{\nu]}=0$ the components  $\Delta_t$, $\Delta_x$, and $\Delta_{+}$ do not mix with the component $\Delta_{-}$ since these have different parities and the background is invariant under parity transformation. Thus, in order to find the general solution one can first ignore the component $\Delta_{-}$ and integrate for $\Delta_t$, $\Delta_x$, and $\Delta_{+}$; then, set $\Delta_t$, $\Delta_x$, and $\Delta_{+}$ to zero and find $\Delta_{-}$. This separation represents no loss of generality.

Finally, suppose that we want integrate the gravitational perturbation equation given in Eq. (\ref{Eqh}) for the case $d=2$. Following the idea presented in the previous paragraph for the gauge 1-form, it follows that the components $h_{tt}$, $h_{xx}$, and $h_{tx}$ should be expanded in terms of spherical harmonics, while the components $h_{ta_2}$ and $h_{xa_2}$ are components of 1-forms with respect to the sphere and, therefore, should be expanded in terms of $V^{+}_{a_2}$ and $V^{-}_{a_2}$. On the other hand, the components $h_{a_2b_2}$ form a rank two symmetric tensor in the sphere and, therefore, should be expanded in terms of an angular basis that has the same nature. Three options for basis are
\begin{align}
  T^\oplus_{a_2b_2} &= Y_{\ell_2}^{m_2}\, \hat{g}_{a_2b_2} \nonumber \\
  T^+_{a_2b_2} &= \hat{\nabla}_{a_2} \hat{\nabla}_{b_2} Y_{\ell_2}^{m_2} \label{T2} \\
  T^-_{a_2b_2} &= \hat{\epsilon}_{a_2c_2}  \hat{\nabla}_{b_2} \hat{\nabla}^{c_2} Y_{\ell_2}^{m_2} +
\hat{\epsilon}_{b_2c_2}    \hat{\nabla}_{a_2} \hat{\nabla}^{c_2} Y_{\ell_2}^{m_2}\,.\nonumber
\end{align}
Explicit expressions for these tensors are provided in appendix \ref{Sec.Appendix}.
Then, the suitable way to expand $\mathbf{h} = h_{\mu\nu}dx^\mu dx^\nu$ is
\begin{align*}
 \mathbf{h} = & e^{i\omega t}\,\big[ ( H_{tt}\, dt^2 + H_{xx} \,dx^2 + 2H_{tx} \, dtdx ) Y_{\ell_2}^{m_2} \\
& 2(H_{t+}dt  +  H_{x+}dx )\,V^{+} +  2(H_{t-}dt  +  H_{x-}dx )\,V^{-} \\
& + H_{\oplus}\,T^\oplus + H_{+}\,T^+ +  H_{-}\,T^-\,\big]\,,
\end{align*}
where $T^+$ stands for $T^+_{a_2b_2}dx^{a_2}dx^{b_2}$, and likewise for $T^\oplus$ and $T^-$. The ten $H$'s are functions only of $x$ and they account for the ten degrees of freedom associated to $h_{\mu\nu}$ in four dimensions. $T^\oplus$ and $T^+$ have even parity, namely transform in the same way as the scalar $Y_{\ell_2}^{m_2}$ under a parity transformation, while $T^-$ has odd parity. With this ansatz for $h_{\mu\nu}$, it is much easier to integrate Eq. (\ref{Eqh}) than using just the spherical harmonics to expand the angular part of the field.

With these examples at hand, we are ready to expand, in a natural way, the gravitational perturbation $\mathbf{h} = h_{\mu\nu}dx^\mu dx^\nu$ in the generalized Nariai spacetime for arbitrary $d$. The components $h_{tt}$,  $h_{xx}$, and $h_{tx}$ are scalars with respect to the $(d-1)$ spheres and, therefore, their angular dependence should be given by the product of spherical harmonics
\begin{equation*}
  \mathcal{Y}= Y_{\ell_2}^{m_2}(\theta_2,\phi_2)\, Y_{\ell_3}^{m_3}(\theta_3,\phi_3)\,\cdots \, Y_{\ell_d}^{m_d}(\theta_d,\phi_d)\,.
\end{equation*}
When we perform a parity transformation in each of the spheres this scalar transforms as
\begin{equation*}
  \mathcal{Y} \xrightarrow{\textrm{parity}} (-1)^{\ell_2 + \cdots + \ell_d} \,\mathcal{Y} \,.
\end{equation*}
Objects that transform in the same way as $\mathcal{Y}$ under a parity transformation will be said to have even parity, while objects that gains an extra minus sign, compared to $\mathcal{Y}$, will be said to have odd parity.

In turn, $h_{ta_j}$ and $h_{xa_j}$ behave as the components of a 1-form with respect to rotations in the $j$-th sphere, but behave as scalars with respect to rotations in the other $(d-2)$ spheres. Thus, a suitable basis for their angular dependence would be
\begin{equation*}
  \mathcal{V}_{a_j}^{\pm} = V^{\pm}_{a_j}(\theta_j,\phi_j)  \prod_{n=2,n\neq j }^d\,  Y_{\ell_n}^{m_n}(\theta_n,\phi_n)\,,
\end{equation*}
where $ V^{\pm}_{a_j}$ have been defined in Eq. (\ref{V2}). With these objects we can define the 1-forms
\begin{equation*}
  \mathcal{V}_{j}^{\pm} = \mathcal{V}_{\theta_j}^{\pm}\, d\theta_j + \mathcal{V}_{\phi_j}^{\pm} \, d\phi_j \,.
\end{equation*}
One can check that the 1-form $\mathcal{V}_{j}^{+}$ has even parity, while $\mathcal{V}_{j}^{-}$ has odd parity.

In an analogous fashion, $h_{a_jb_j}$ behave as a symmetric rank two tensor with respect to rotations in the $j$-th sphere and as scalars with respect to rotations in the $n$-th sphere when $n\neq j$. Thus, a suitable basis for the angular dependence of this part is
\begin{equation*}
  \mathcal{T}^{\pm,\oplus}_{a_j b_j} = T^{\pm,\oplus}_{a_jb_j}(\theta_j,\phi_j)  \prod_{n=2,n\neq j }^d\,  Y_{\ell_n}^{m_n}(\theta_n,\phi_n)\,,
\end{equation*}
where $T^{+}_{a_jb_j}$, $T^{-}_{a_jb_j}$, and $T^{\oplus}_{a_jb_j}$ have been defined in Eq. (\ref{T2}).
The corresponding tensors are, then, defined by
\begin{equation*}
  \mathcal{T}^{\pm,\oplus}_{j} = \mathcal{T}^{\pm,\oplus}_{\theta_j \theta_j} d\theta_j^2 \,+\,
 2\mathcal{T}^{\pm,\oplus}_{\theta_j \phi_j} d\theta_j d\phi_j
\,+\, \mathcal{T}^{\pm,\oplus}_{\phi_j \phi_j} d\phi_j^2 \,,
\end{equation*}
where $\mathcal{T}^{\oplus}_{j}$ and $\mathcal{T}^{+}_{j}$ have even parity, while $\mathcal{T}^{-}_{j}$ has odd parity.

A more tricky type of component is $h_{a_nb_j}$ with $n\neq j$, which behaves as the components of a 1-form under rotations in the $n$-th and $j$-th
spheres, while it behaves as a scalar with respect to rotations in the other spheres. We need a basis for the angular dependence that has this property. On top of that, we would like the basis elements to have a definite parity. A solution to these constraints is provided by the following functions:
\begin{align*}
   \mathcal{W}^{+}_{a_j b_n} &= V^{+}_{a_j}(\theta_n,\phi_n)  V^{+}_{b_n}(\theta_j,\phi_j)  \prod_{k\neq n,j }   Y_{\ell_k}^{m_k}(\theta_k,\phi_k) \,,\\
\mathcal{W}^{\oplus}_{a_j b_n} &= V^{-}_{a_j}(\theta_i,\phi_i)  V^{-}_{b_n}(\theta_j,\phi_j)
 \prod_{k\neq n,j }   Y_{\ell_k}^{m_k}(\theta_k,\phi_k) \,, \\
 \mathcal{W}^{-}_{a_j b_n} &= V^{+}_{a_j}(\theta_i,\phi_i)  V^{-}_{b_n}(\theta_j,\phi_j)
\prod_{k\neq n,j }   Y_{\ell_k}^{m_k}(\theta_k,\phi_k) \,,\\
\mathcal{W}^{\ominus}_{a_j b_n} &= V^{-}_{a_j}(\theta_i,\phi_i)  V^{+}_{b_n}(\theta_j,\phi_j)
\prod_{k\neq n,j }   Y_{\ell_k}^{m_k}(\theta_k,\phi_k) \,,
\end{align*}
Using  these components we can define the symmetric rank two tensors $\mathcal{W}_{jn}^{+}$,  $\mathcal{W}_{jn}^{\oplus}$, $\mathcal{W}_{jn}^{-}$, and $\mathcal{W}_{jn}^{\ominus}$ in the natural way. For instance,
\begin{align*}
  \mathcal{W}_{jn}^{+} =\, & \mathcal{W}^{+}_{\theta_j \theta_n} d\theta_j d\theta_n \,+\,
  \mathcal{W}^{+}_{\theta_j \phi_n} d\theta_j d\phi_n \\
&\,+\, \mathcal{W}^{+}_{\phi_j \theta_n} d\phi_j d\theta_n \,+\, \mathcal{W}^{+}_{\phi_j \phi_n} d\phi_j d\phi_n   \,,
\end{align*}
and analogously for the other three tensors. $\mathcal{W}^{+}$ and $\mathcal{W}^{\oplus}$ have positive parity, as they transform in the same way as $\mathcal{Y}$ under a parity transformation, while $\mathcal{W}^{-}$ and $\mathcal{W}^{\ominus}$ have negative parity. Note that the odd parity modes come from the product of modes with opposite parities, whereas the positive parity modes arise from the product of elements with the same parity. Moreover, note that in order to consider $\mathcal{W}_{jn}^{+}$, $\mathcal{W}_{jn}^{\oplus}$, $\mathcal{W}_{jn}^{-}$, and $\mathcal{W}_{jn}^{\ominus}$ as a basis for the components of the type $h_{a_jb_n}$, with $n\neq j$, we just need to assume $n>j$, since the case $j<n$ lead to the same rank two tensors. For instance, $\mathcal{W}^{+}_{23} = \mathcal{W}^{+}_{32} $, and $\mathcal{W}^{-}_{23} = \mathcal{W}^{\ominus}_{32}$.

The preceding steps used to find a suitable basis for the angular dependence should not be underestimated. Indeed, the perturbation equation for $h_{\mu\nu}$ is quite involved and can lead to an unbearable entanglement between the components of $h_{\mu\nu}$  if a natural basis is not adopted.

Thus, using these angular bases, a suitable way to expand the gravitational perturbation $\mathbf{h}=h_{\mu\nu}dx^\mu dx^\nu$ is as follows:
\begin{align*}
  \mathbf{h}&= e^{i\omega t}\Big[   (H_{tt} dt^2 + 2 H_{tx} dtdx + H_{xx} dx^2) \mathcal{Y}   \\
& + \sum_{j=2}^d  \left( H_{tj}^+ \mathcal{V}_{j}^{+} +  H_{tj}^- \mathcal{V}_{j}^{-} \right) dt +
\left( H_{xj}^+ \mathcal{V}_{j}^{+} +  H_{xj}^- \mathcal{V}_{j}^{-} \right) dx\\
& + \sum_{j=2}^d  H_j^+ \,\mathcal{T}_{j}^{+} +  H_j^\oplus \,\mathcal{T}_{j}^{\oplus} + H_j^- \,\mathcal{T}_{j}^{-}  \\
& + \sum_{j=2}^d \sum_{n>j}^d  H_{jn}^+  \mathcal{W}_{jn}^{+} +  H_{jn}^\oplus  \mathcal{W}_{jn}^{\oplus}
+ H_{jn}^-  \mathcal{W}_{jn}^{-} + H_{jn}^\ominus  \mathcal{W}_{jn}^{\ominus} \Big],
\end{align*}
where the $H$'s are all functions of the coordinate $x$. Counting the number of independent functions, we have three coming from the first line of the right hand side of the previous equation, namely from $H_{tt}$,  $H_{tx}$, and $H_{xx}$; in the second line there are $(d-1)$ functions  $H_{tj}^+$ and, analogously, more $3(d-1)$ components stemming from $H_{tj}^-$, $H_{xj}^+$, and $H_{xj}^-$; in the third line we have $3(d-1)$ independent functions; finally, in the fourth line we should recall that $n>j$, so that there are $4\frac{(d-1)(d-2)}{2}$ functions. Summing the number of these functions, we have:
\begin{equation*}
  3 + 4(d-1) + 3(d-1) + 2(d-1)(d-2) = \frac{2d(2d+1)}{2},
\end{equation*}
which is exactly the number of independent components of $h_{\mu\nu}$ in $2d$ dimensions. This proves that no possible degree of freedom of the perturbation field is being neglected.

Once we have made an appropriate expansion for $h_{\mu\nu}$, we are ready to start the integration process of Eq. (\ref{Eqh}). In order to do so, we can take advantage of the spherical symmetries and only consider the cases $m_j = 0$, for all $j$, so that no $\phi_j$ dependence will show up. Thus, any derivative of the type $\partial_{\phi_j}$ will not contribute, including those appearing in the definition of our basis. For instance, $\mathcal{V}_{\phi_j}^{+}$ and $\mathcal{V}_{\theta_j}^{-}$ are automatically zero in such a case (see the expressions in appendix \ref{Sec.Appendix}). As explained before, this will represent no important loss of generality, since the other solutions can be generated by applying rotations to the ones with $m_j=0$. Moreover, in this work we are only interested in the frequencies of the quasinormal modes, which are invariant under the rotations in the spheres, so that we do not even need to bother about generating solutions with nonzero values of $m_j$. Another great simplification that we can take advantage of stems from the fact that the field equation for $h_{\mu\nu}$ do not mix components with opposite parities.  Thus, in what follows we will separate the integration of the perturbation equation in the odd degrees of freedom, which will be tackled in the next section, and the even degrees of freedom, which will be considered  section \ref{Sec.Even}.

A different source of simplification in the calculations performed below arise from the gauge freedom in the choice of the coordinate system. If we perform
the change in the coordinates
\begin{equation}\label{CoordTransf}
  x^\mu \,\mapsto\, \tilde{x}^\mu = x^\mu +  \zeta^\mu \,,
\end{equation}
where $\zeta^\mu = \zeta^\mu(x)$ is infinitesimal, it follows that the components of the metric in the new coordinate system are given by
\begin{equation*}
  g_{\mu\nu} \,\mapsto\,  \tilde{g}_{\mu\nu} = g_{\mu\nu} + \nabla_\mu \zeta_\nu +  \nabla_\nu \zeta_\mu \,.
\end{equation*}
Thus, performing the perturbation (\ref{PerturbPhi}) in the metric  followed by the infinitesimal coordinate transformation (\ref{CoordTransf}) is equivalent, to first order in the infinitesimal parameters, to performing just a metric perturbation with the perturbation field being
\begin{equation*}
  \tilde{h}_{\mu\nu} = h_{\mu\nu} + \nabla_\mu \zeta_\nu +  \nabla_\nu \zeta_\mu \,.
\end{equation*}
Since physics is insensitive to coordinate transformations, it follows that the transformation
\begin{equation}\label{GaugetTransf}
  h_{\mu\nu}\,\mapsto\,  h_{\mu\nu} + \nabla_\mu \zeta_\nu +  \nabla_\nu \zeta_\mu \,.
\end{equation}
is just a gauge transformation, namely it does not lead to changes in the physical results. In particular, these transformations do not change the quasinormal spectrum of the gravitational perturbation (assuming that we are not changing our time component). In what follows we will perform a wise choice for the vector field $\zeta^\mu$ in order to eliminate some degrees of freedom of the perturbation field.

\section{Odd Perturbations}\label{Sec.Odd}

Constraining the perturbation to have just the odd degrees of freedom, under the parity transformation, it follows that we can assume that $h_{\mu\nu}$ is given by:
\begin{align}
  \mathbf{h}= e^{i\omega t}\Big\{&    \sum_{j=2}^d  \left[ \left(  H_{tj}^- \,dt + H_{xj}^- \,dx \right) \mathcal{V}_{j}^{-} \,+\, H_j^- \,\mathcal{T}_{j}^{-}  \right] \nonumber\\
& + \sum_{j=2}^d \sum_{n>j}^d \left(  H_{jn}^- \,\mathcal{W}_{jn}^{-} + H_{jn}^\ominus \,\mathcal{W}_{jn}^{\ominus} \right) \Big\}\,.\label{hodd}
\end{align}
However, one can eliminate some degrees of freedom by means of a gauge transformation. Indeed, performing the transformation (\ref{GaugetTransf}) with $\zeta_\mu $ given by
\begin{equation*}
  \zeta_\mu dx^\mu = - e^{i\omega t}\sum_{j=2}^d H_j^-\,\mathcal{V}_{j}^{-} \,,
\end{equation*}
it follows that the transformed field $\tilde{h}_{\mu\nu}$ is such that it has the same form as depicted in the expansion (\ref{hodd}) but with the fields $H^-(x)$ transformed to $\tilde{H}^-(x)$ where
\begin{equation*}
\left\{
  \begin{array}{ll}
     \tilde{H}_{tj}^- =  H_{tj}^- - i\omega H_j^- \,, \\
    \tilde{H}_{xj}^- =  H_{xj}^- - \frac{d}{dx} H_j^- \,,\\
    \tilde{H}_{j}^- = 0 \,,\\
     \tilde{H}_{jn}^- =  H_{jn}^- -  H_n^- \,,\\
     \tilde{H}_{jn}^\ominus =  H_{jn}^\ominus -   H_j^- \,.
  \end{array}
\right.
\end{equation*}
Thus, we see that the components  $H_{j}^-$ of the ansatz (\ref{hodd}) can be eliminated by a gauge transformation, while the other components just get redefined. Thus, in what follows we can ignore the degrees of freedom $\tilde{H}_{j}^-$ and consider that the gravitational perturbation is given by
\begin{align*}
  \mathbf{h}= e^{i\omega t}\Big\{&    \sum_{j=2}^d  \left[ \left(  H_{tj}^- \,dt + H_{xj}^- \,dx \right) \mathcal{V}_{j}^{-}   \right] \nonumber\\
& + \sum_{j=2}^d \sum_{n>j}^d \left(  H_{jn}^- \,\mathcal{W}_{jn}^{-} + H_{jn}^\ominus \,\mathcal{W}_{jn}^{\ominus} \right) \Big\}\,.\label{hodd2}
\end{align*}

Now, inserting this perturbation into the field equation (\ref{Eqh}), we are eventually led to the following equations:
\begin{widetext}
\begin{equation}\label{Emenos}
  \left.
     \begin{array}{ll}
      E^-_{t\phi_j}\,\equiv & \, \frac{d}{dx}\left[ \frac{1}{f}\left( \frac{d}{dx} H^-_{tj} - i \omega  H^-_{xj} \right) \right] -
\Lambda  \,(\kappa -2  ) \,H^-_{tj} + i\omega \Lambda \sum_{n\neq j} \kappa_n  ( H_{jn}^\ominus + H_{nj}^-)  = 0 \,,  \\
\\
E^-_{x\phi_j} \,\equiv & \,  \frac{i\omega}{f}\left( \frac{d}{dx} H^-_{tj} - i \omega  H^-_{xj} \right) -
\Lambda \, (\kappa - 2) \,H^-_{xj} +   \Lambda \sum_{n\neq j} \kappa_n \frac{d}{dx}\left( H_{jn}^\ominus + H_{nj}^-\right)  = 0  \,, \\
\\
E^-_{\theta_j\phi_j} \,\equiv & \,  \frac{1}{f}\left( \frac{d}{dx} H^-_{xj} - i \omega  H^-_{tj} \right) -
  \Lambda \sum_{n\neq j} \kappa_n \left(  H_{jn}^\ominus + H_{nj}^-\right)  = 0  \,,  \\
\\
  E^-_{\theta_j\phi_n} \,\equiv & \, \frac{d^2}{dx^2}\left(  H_{nj}^\ominus + H_{jn}^-\right) +
  \left[\omega^2 - f\,\Lambda\, (\kappa - 2) \right]\left(  H_{nj}^\ominus + H_{jn}^-\right)
  -f\,E^-_{\theta_n\phi_n} = 0  \,.
     \end{array}
   \right.
\end{equation}
\end{widetext}
In the left hand side of these equations, the objects $E^-_{\mu\nu}$ are just to stress that the equation $E^-_{\mu\nu}=0$ comes from imposing the component $\mu\nu$ of Eq. (\ref{Eqh}) to hold. The components that do not appear above, like $E^-_{tt}$ are identically vanishing. In the last line of Eq. (\ref{Emenos}) it is being assumed that $n\neq j$.  Above, we have also used the definitions
\begin{equation}
 \kappa_j = \ell_j(\ell_j + 1) \;\; \textrm{and} \;\;   \kappa =\sum_{j=2}^d\,\kappa_j\,.
\end{equation}
Thus, the above equations comprise all the restrictions associated to the perturbation equation obeyed by $h_{\mu\nu}$.

In order to attain Eq. (\ref{Emenos}), we have assumed that the spherical harmonics $Y_{\ell_j}^{m_j}(\theta_j,\phi_j)$ have $m_j=0$, which is justified by the spherical symmetry, as explained before. So, we have used $Y_{\ell_j}^{m_j} = Y_{\ell_j}(\theta_j)$ where $Y_{\ell_j}(\theta_j)$ obeys the following differential equation:
\begin{equation*}
  \frac{1}{\sin\theta_j}\frac{d}{d\theta_j}\left( \sin\theta_j\,\frac{d}{d\theta_j} Y_{\ell_j} \right) + \kappa_j Y_{\ell_j} = 0\,.
\end{equation*}

In the equations displayed in (\ref{Emenos}) , the fields $H_{jn}^\ominus$ and $H_{jn}^-$ appear only by means of the combination $(H_{nj}^\ominus + H_{jn}^-)$. Note, however, that we are always assuming that $n\neq j$, so that either $n>j$ or $n<j$. These fields were defined through Eq. (\ref{hodd}), where it is always assumed that the second index is greater than the first. Thus, the fields  $H_{jn}^\ominus$ and $H_{jn}^-$ with $j>n$ are not defined. Hence, the convention in Eq. (\ref{Emenos}) is that these undefined fields are zero. So, what might appear as two fields in the sum $(H_{jn}^\ominus + H_{nj}^-)$ is, actually, just one field. Indeed, if $n>j$ it follows that $H_{nj}^-$ vanishes, so that $(H_{jn}^\ominus + H_{nj}^-) = H_{jn}^\ominus$, while if $j>n$ we have $(H_{jn}^\ominus + H_{nj}^-) = H_{nj}^-$. Summing up, in Eq. (\ref{Emenos}) we have
\begin{equation*}
  (H_{jn}^\ominus + H_{nj}^-) \,=\, \left\{
                                      \begin{array}{ll}
                                        H_{nj}^- \;,\;\; \textrm{if } j> n\\
                                        \;\\
                                        H_{jn}^\ominus \;,\;\; \textrm{if } n>j
                                      \end{array}
                                    \right.\,.
\end{equation*}

\begin{figure}[h]
  \centering
  \includegraphics[width=7.5cm]{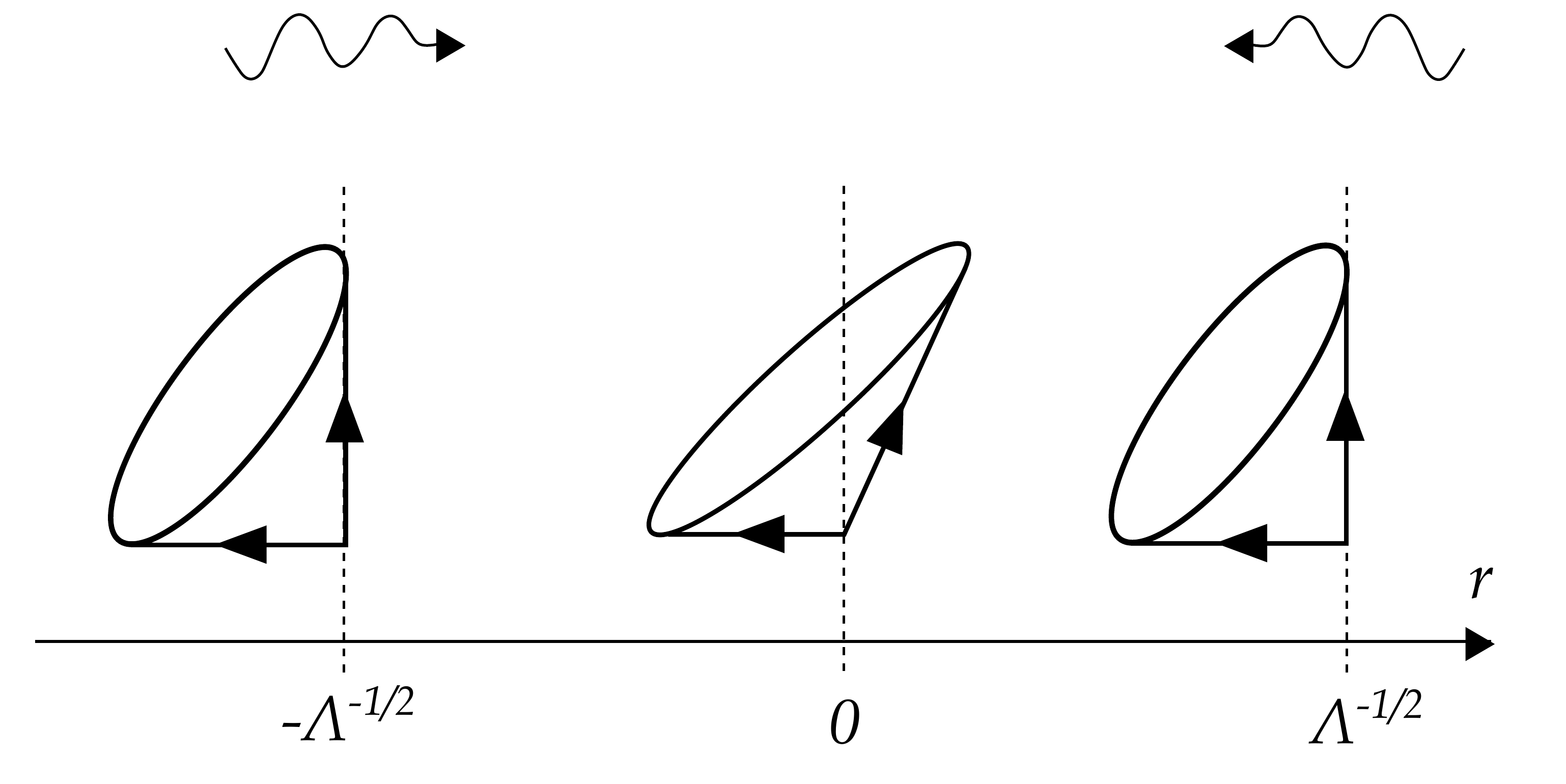}
  \caption{The wavy arrows depicts the direction of the perturbation field at the boundaries, while the cones are the local light cones. This boundary conditions is the appropriate one to attain a quasinormal mode at the generalized Nariai spacetime, see Ref. \cite{JoasI}. }
  \label{FigCones10}
\end{figure}

Assuming that $E^-_{\theta_j\phi_j}$ vanishes, in accordance with the third equation in (\ref{Emenos}), it follows from the last line in (\ref{Emenos}) that the fields $H_{jn}^-$ and $ H_{nj}^\ominus$ both obey the following differential equation:
\begin{equation}\label{PoschlTeller}
  \frac{d^2}{dx^2}H + \left[ \omega^2 - \frac{\Lambda\, (\kappa - 2) }{\cosh^2(\sqrt{\Lambda}\,x )}\right] H = 0 \,.
\end{equation}
This is the well-known P\"{o}schl-Teller equation \cite{Teller}, that can be integrated analytically. In particular, assuming that the boundary condition for the perturbation field is as depicted in Fig. \ref{FigCones10}, which is the appropriate boundary condition for quasinormal modes, it follows that the spectrum of allowed frequencies is
\begin{equation}\label{Spectrum}
  \omega = \sqrt{\Lambda}\left[\, \sqrt{\kappa -9/4 } \,+\, i \,\left(N+1/2\right) \,\right] \,,
\end{equation}
where $N\in\{0,1,2,\cdots \}$. For more details on the calculation of the spectrum and on the choice of boundary condition, the reader is referred to Ref. \cite{JoasI}.
Thus, summing up, we have just proved that the spectrum of the degrees of freedom $H_{jn}^-$ and $ H_{jn}^\ominus$  is the one given in Eq. (\ref{Spectrum}). It remains to check whether $H^-_{tj}$ and $H^-_{xj}$ have the same spectrum. Defining the field
\begin{equation*}
  \breve{H}^-_j =\frac{1}{f}\left( \frac{d}{dx} H^-_{tj} - i \omega  H^-_{xj} \right) \,,
\end{equation*}
it follows immediately from the equation
$$\frac{d}{dx}E_{t\phi_j} - i\omega E_{x\phi_j} = 0$$
that $\breve{H}^-_j$ also obeys the P\"{o}schl-Teller equation (\ref{PoschlTeller}) and, therefore, have the same spectrum of the fields $H_{jn}^-$ and $ H_{jn}^\ominus$, namely (\ref{Spectrum}). Then, by means of the equations $E^-_{t\phi_j}=0$ and $E^-_{x\phi_j}=0$ we can write the fields $H^-_{tj}$ and $H^-_{xj}$ in terms of the fields that obey the P\"{o}schl-Teller equation. More precisely, we have:
\begin{equation*}
     \begin{array}{ll}
       H^-_{tj} &= \frac{1}{\Lambda(\kappa - 2)} \frac{d}{dx} \breve{H}^-_j  + \frac{i\omega}{\kappa - 2}\sum_{n\neq j} \kappa_n ( H_{jn}^\ominus + H_{nj}^-)\,,\\
\\
       H^-_{xj} &= \frac{i\omega}{\Lambda(\kappa - 2)} \breve{H}^-_j + \frac{1}{\kappa - 2}\sum_{n\neq j} \kappa_n \frac{d}{dx}( H_{jn}^\ominus + H_{nj}^-)\,.
     \end{array}
\end{equation*}
So, $H^-_{tj}$ and $H^-_{xj}$  must have the same spectrum of $\breve{H}^-_j$, $H^-_{jn}$, and $H^\ominus_{jn}$, namely (\ref{Spectrum}). Indeed, since fields $\breve{H}^-_j$, $H^-_{jn}$, and $H^\ominus_{jn}$ obey the boundary condition depicted in Fig. \ref{FigCones10}, it follows that near the boundaries $x\rightarrow \pm \infty$ ($r\rightarrow \pm 1/\sqrt{\Lambda}$) the behaviour of these fields is $e^{\pm i \omega x}$. Thus, linear combinations of these fields and their derivatives will also obey the same boundary conditions. Another way to understand why $H^-_{tj}$ and $H^-_{xj}$ have the spectrum (\ref{Spectrum}) is by applying the differential operator that acts on $H$ in Eq. (\ref{PoschlTeller}) to the above expressions for $H^-_{tj}$ and $H^-_{xj}$. Doing so, we can check that $H^-_{tj}$ and $H^-_{xj}$ obey the P\"{o}schl-Teller equation with a source, namely
\begin{equation*}
   \left[ \frac{d^2}{dx^2}+ \omega^2 - \frac{\Lambda\, (\kappa - 2) }{\cosh^2(\sqrt{\Lambda}\,x )}\right] H^-_{tj} =
F_j   \frac{df}{dx} \,,
\end{equation*}
where $F_j=F_j(x)$ is some field obeying the P\"{o}schl-Teller equation and likewise for  $H^-_{xj}$. The general solution for a linear differential equation with a source is given by the general solution for the homogeneous part of the equation, which in the latter case is the P\"{o}schl-Teller equation, plus a particular solution that depends linearly on the source. In the case of interest, the source goes to exponentially zero at the boundaries, due to the term $df/dx$. Hence, near the boundaries $H^-_{tj}$ and $H^-_{xj}$ obey the P\"{o}schl-Teller equation and, therefore, yield the same spectrum (\ref{Spectrum}).

So far, we have imposed and solved the equations $ E^-_{t\phi_j}=0$, $ E^-_{x\phi_j}=0$, and $ E^-_{\theta_j\phi_n}=0$, whereas we have just assumed $ E^-_{\theta_j\phi_j}=0$ to be true, without really solving it. However, inserting the latter expressions for $H^-_{tj}$ and $H^-_{xj}$ in the third line of Eq. (\ref{Emenos}) it follows that $ E^-_{\theta_j\phi_j}=0$ whenever $H_{jn}^\ominus$ and $H_{jn}^-$ obey the P\"{o}schl-Teller equation (\ref{PoschlTeller}), so that the constraint $ E^-_{\theta_j\phi_j}=0$ is already guaranteed to hold once the other equations in (\ref{Emenos}) are solved.  In conclusion, all degrees of freedom of the odd perturbation have the spectrum (\ref{Spectrum}).

\section{Even Perturbations}\label{Sec.Even}

The even parity perturbation has the general form
\begin{align}
  \mathbf{h}&= e^{i\omega t}\Big[   (H_{tt} dt^2 + 2 H_{tx} dtdx + H_{xx} dx^2) \mathcal{Y}   \nonumber \\
& + \sum_{j=2}^d  \left( H_{tj}^+ \mathcal{V}_{j}^{+} dt + H_{xj}^+ \mathcal{V}_{j}^{+} dx + H_j^+ \,\mathcal{T}_{j}^{+} +  H_j^\oplus \,\mathcal{T}_{j}^{\oplus} \right) \nonumber \\
& + \sum_{j=2}^d \sum_{n>j}^d  H_{jn}^+ \,\mathcal{W}_{jn}^{+} +  H_{jn}^\oplus \,\mathcal{W}_{jn}^{\oplus}\Big]\,, \label{heven1}
\end{align}
Then, performing a gauge transformation (\ref{GaugetTransf}) with
\begin{equation*}
  \zeta_\mu dx^\mu =\frac{ e^{i\omega t}}{2} \left[ A \,\mathcal{Y}dt + B \,\mathcal{Y} dx - \sum_{j=2}^d H_{j}^{+}\,\mathcal{V}_{j}^{+} \right] \,,
\end{equation*}
where $A = A(x)$ and $B = B(x)$ are functions of the coordinate $x$ given by:
\begin{eqnarray}
A & = & i \omega\,H_{j}^{+}\,-\,2 H_{t2}^{+}\,,\\
B & = & \frac{d}{dx}\,H_{j}^{+}\,-\,2 H_{x2}^{+} \,,
\end{eqnarray}
it follows that the transformed perturbation field $\tilde{h}_{\mu\nu}$ is such that it admits an expansion just as depicted in Eq. (\ref{heven1}) but with the fields $H(x)$ transformed to $\tilde{H}(x)$ where $\tilde{H}_{j}^{+}$, $\tilde{H}_{t2}^{+}$, and $\tilde{H}_{x2}^{+}$ all vanish; $\tilde{H}_{j}^{\oplus}$ and $\tilde{H}_{jn}^{\oplus}$ are equal to $H_{j}^{\oplus}$ and $H_{jn}^{\oplus}$ respectively; while the other degrees of freedom change as follows:
\begin{widetext}
\begin{eqnarray}
\tilde{H}_{tt} &=& H_{tt}\,-\,\frac{f'}{2f}\left(\frac{d}{dx}H_{2}^{+}\,-\,2H_{x2}^{+} \right ) \,-\,\omega^{2}H_{2}^{+}\,-\,2i\omega H_{t2}^{+} \,,\nonumber\\
\tilde{H}_{xx} &=& H_{xx}\,-\,\frac{f'}{2f}\left(\frac{d}{dx}H_{2}^{+}\,-\,2H_{x2}^{+} \right ) \,+\,\frac{d^{2}}{dx^{2}}H_{2}^{+}\,-\,2\frac{d}{dx}H_{x2}^{+} \,,\nonumber\\
\tilde{H}_{tx}&=& H_{tx}\,-\,\frac{f'}{2f}\left( i \omega H_{2}^{+}\,-\,2H_{t2}^{+} \right ) \,+\,i\omega \frac{d}{dx}H_{2}^{+}\,-\,\frac{d}{dx}H_{t2}^{+}\,-\,i\omega H_{x2}^{+} \,,\nonumber\\
\tilde{H}_{tj}^{+} &=& H_{tj}^{+}\,-\,H_{t2}^{+}\,+\,\frac{i\omega}{2}\left(H_{2}^{+}\,-\,H_{j}^{+}\right )  \quad \forall \, j\neq 2 \,,\nonumber\\
\tilde{H}_{xj}^{+} &=& H_{xj}^{+}\,-\,H_{x2}^{+}\,+\,\frac{1}{2}\frac{d}{dx}\left(H_{2}^{+}\,-\,H_{j}^{+}\right ) \quad \forall \, j\neq 2 \,,\nonumber\\
\tilde{H}_{jn}^{+} &=& H_{jn}^{+}\,-\,\frac{1}{2}\left(H_{j}^{+}\,+\,H_{n}^{+} \right )\,.\nonumber
\end{eqnarray}
\end{widetext}
Hence, the following $(d+1)$ degrees of freedom can be set to zero:
\begin{equation}
\tilde{H}_{t2}^{+} \,=\, 0 \quad , \quad \tilde{H}_{x2}^{+} \,=\, 0 \quad , \quad \tilde{H}_{j}^{+} \,=\, 0 \,,
\end{equation}
since they can be eliminated by a gauge transformation, whereas the other $\tilde{H}$'s are just equal to the previous $H$'s added by functions of $x$. Thus, assuming this gauge choice and dropping the tildes, it follows that we can assume that the perturbation field has the form
\begin{align}
  \mathbf{h}&= e^{i\omega t}\Big[   (H_{tt} dt^2 + 2 H_{tx} dtdx + H_{xx} dx^2) \mathcal{Y} \nonumber \\
  &\quad + \sum_{j=2}^d H_{j}^\oplus \,\mathcal{T}_{j}^{\oplus}
   + \sum_{j=3}^d  \left( H_{tj}^+ \mathcal{V}_{j}^{+} dt + H_{xj}^+ \mathcal{V}_{j}^{+} dx \right) \nonumber \\
& + \sum_{j=2}^d \sum_{n>j}^d \left(H_{jn}^+ \,\mathcal{W}_{jn}^{+} +  H_{jn}^\oplus \,\mathcal{W}_{jn}^{\oplus}\right)\Big]\,, \label{heven2}
\end{align}

Now, inserting this ansatz into the field equation \eqref{Eqh}, we find the following differential equations obeyed by the fields $H$'s
\begin{widetext}
\begin{eqnarray}\label{Emais}
E^+_{tt} &\equiv&  \Lambda\,(2 - \kappa f)\,H_{tt} - 2\Lambda\,H_{xx} + 2\Lambda f \sum_{j=2}^{d}\left[\,i \omega \left(\kappa_{j}\,H_{tj}^{+} + i\omega H_j^{\oplus} \right) - \frac{f'}{2f}\left(\kappa_{j} H_{xj}^{+} + \frac{d}{dx}H_j^{\oplus}\right ) \right] +\nonumber\\
&+& \frac{d^2}{dx^2} H_{tt} - 2i\omega\, \frac{d}{dx}H_{tx} - \omega^{2}H_{xx} + \frac{f'}{2f}\left(\frac{d}{dx} H_{xx} -3\,\frac{d}{dx}H_{tt} + 2i\omega\, H_{tx} \right ) \,=\,0 \,,\nonumber\\
E^+_{tx} &\equiv&  \sum_{j}\left[ \kappa_j f \frac{d}{dx}\left( \frac{1}{f}H_{tj}^+ \right) + i\omega \kappa_j H_{xj}^+
- \kappa_j\, H_{tx} + 2i\omega \sqrt{f}\frac{d}{dx}\left( \frac{1}{\sqrt{f}}H_j^{\oplus} \right)   \right] = 0  \,, \nonumber\\
E^+_{xx} &\equiv&  2\Lambda\,H_{tt} - \Lambda\,(2 - \kappa f)\,H_{xx} - 2\Lambda f \sum_{j=2}^{d}\left[\kappa_{j}\,\frac{d}{dx} H_{xj}^{+} + \frac{d^2}{dx^2} H_j^{\oplus} - \frac{f'}{2f}\left(\kappa_{j} H_{xj}^{+} + \frac{d}{dx} H_j^{\oplus}\right ) \right] +\nonumber\\
&+& \frac{d^2}{dx^2} H_{tt} - 2i\omega\, \frac{d}{dx} H_{tx} - \omega^{2}H_{xx} + \frac{f'}{2f}\left(\frac{d}{dx} H_{xx} -3\,\frac{d}{dx}H_{tt} + 2i\omega\, H_{tx} \right ) \,=\,0 \,,\nonumber\\
E^+_{t\theta_j} &\equiv& \frac{d}{dx}\left[\frac{1}{f}\left(\frac{d}{dx}H_{tj}^+ - i \omega H_{xj}^+  \right) \right] + i\omega \left[\frac{1}{f}\left(\frac{d}{dx}H_{xj}^{+} - i \omega H_{tj}^+  \right) \right]
- \Lambda (\kappa-2) H_{tj}^{+} \nonumber\\
&-& \frac{1}{f}\,\frac{d}{dx} H_{tx} + \frac{i \omega}{2f}\,(H_{tt} + H_{xx}) + \Lambda \sum_{n=2}^{d}\left(i\omega H_n^{\oplus} + \kappa_{n} H_{tn}^{+}  \right) - i \omega  E^{I}_{\phi_{j}\phi_{j}} = 0 \,,\nonumber\\
E^+_{x\theta_j} &\equiv& i\omega \left[\frac{1}{f}\left(\frac{d}{dx}H_{tj}^+ - i \omega H_{xj}^+  \right) \right] + \frac{d}{dx} \left[\frac{1}{f}\left(\frac{d}{dx}H_{xj}^{+} - i \omega H_{tj}^+  \right) \right]
- \Lambda (\kappa-2) H_{xj}^{+} +\nonumber\\
&+& \frac{i\omega}{f}\,H_{tx} -\frac{1}{2f}\,\frac{d}{dx}\,(H_{tt} + H_{xx}) + \Lambda \sum_{n=2}^{d}\left(\frac{d}{dx}H_n^{\oplus} + \kappa_{n} H_{xn}^{+}  \right) - \frac{d}{dx}  E^{I}_{\phi_{j}\phi_{j}} = 0 \,,\nonumber\\
E^+_{\theta_j\theta_n} &\equiv& \frac{d^2}{dx^2} (H_{jn}^+ + H_{nj}^+ ) + \left[ \omega^2 - \Lambda f (\kappa-2) \right] (H_{jn}^+ + H_{nj}^+ )
- f\, E^{I}_{\phi_j\phi_j} - f\,E^{I}_{\phi_n\phi_n} = 0 \,,\nonumber\\
E^+_{\phi_j\phi_n} &\equiv&  \frac{d^2}{dx^2} (H_{jn}^\oplus  + H_{nj}^\oplus ) +
 \left[ \omega^2 - \Lambda f (\kappa-2) \right] (H_{jn}^\oplus + H_{nj}^\oplus ) = 0 \,,\nonumber\\
E^{I}_{\phi_j\phi_j} &\equiv&   \frac{1}{f}\left(\frac{d}{dx} H_{xj}^+ - i \omega  H_{tj}^{+} \right) + \Lambda H_{j}^\oplus -
 \Lambda  \sum_{n=2}^{d} \left[  H_{n}^{\oplus}  + \kappa_{n} (H_{jn}^+  + H_{nj}^+ )  \right] + \frac{1}{2f}(H_{tt} - H_{xx}) = 0 \,,\nonumber \\
E^{II}_{\phi_j\phi_j} &\equiv&  \frac{d^2}{dx^2} H_{j}^{\oplus} + \left[ \omega^2 - \Lambda f ( \kappa -2 ) \right] H_{j}^\oplus  = 0 \,, \nonumber\\
E^{I}_{\theta_j\theta_j} &\equiv &  E^{I}_{\phi_j\phi_j} = 0  \,,    \nonumber\\
E^{II}_{\theta_j\theta_j} &\equiv & E^{II}_{\phi_j\phi_j} + 2 f\,\kappa_j \,E^{I}_{\phi_j\phi_j}  = 0 \,,
\end{eqnarray}
\end{widetext}
where $f'$ stands for $df/dx$, as usual.

The great advantage of using the angular basis $\{\mathcal{Y}, \mathcal{V}_{j}^{+},\mathcal{T}_{j}^{\oplus},\cdots \}$, instead of just using the spherical harmonics $\mathcal{Y}$ is that when we compute the components of the perturbation equation (\ref{Eqh}) the angular dependence automatically factors out as a global multiplicative term, so that we end up with equations that depend just on the coordinate $x$, as we have seen for the odd perturbation, in the preceding section, and as we now see in the above equations for the even perturbations. Nevertheless, in the components $\phi_j\phi_j$ and $\theta_j\theta_j$ of the even perturbation equation the angular functions do not factor automatically, rather we face an equation of the following type
\begin{equation}\label{LI}
 Y_{\ell_j}^{0}(\theta_j)\,A(x) + \cot\theta_j \frac{d}{d\theta_j}Y_{\ell_j}^{0}(\theta_j) \,B(x)= 0 \,.
\end{equation}
However, in general, the spherical harmonic $Y_{\ell_j}^{0}$ is linearly independent from $\cot\theta_j \frac{d}{d\theta_j}Y_{\ell_j}^{0}$, so that the latter equation implies both $A(x) =0$ and $B(x)=0$. This is the reason why the equations that stem from the components $\phi_j\phi_j$ and $\theta_j\theta_j$ are split in two separate constraints, which are denoted in Eq. (\ref{Emais}) by $E^{I}_{\phi_j\phi_j}$, $E^{II}_{\phi_j\phi_j}$, $E^{I}_{\theta_j\theta_j}$, and $E^{II}_{\theta_j\theta_j}$. The only case in which we cannot conclude that $A$ and $B$ are both zero in Eq. (\ref{LI}) is when the two angular functions are linearly dependent, namely when
\begin{equation*}
  \alpha \,  Y_{\ell_j}^{0}(\theta_j) + \beta\, \cot\theta_j \frac{d}{d\theta_j}Y_{\ell_j}^{0}(\theta_j) = 0
\end{equation*}
for some constants $\alpha$ and $\beta$. Integrating the latter constraint, we conclude that the linear dependence happens only if
\begin{equation*}
  Y_{\ell_j}^{0}(\theta_j) = c\, (\cos\theta_j)^{\alpha/\beta} \,,
\end{equation*}
where $c$ is some constant. This is true only for $\ell_j=0$, in which case $\alpha/\beta = 0$, and for $\ell_j=1$, in which case $\alpha/\beta = 1$. Thus, for any $\ell_j > 1$ we can promptly assume that, in Eq. (\ref{LI}), $A$ and $B$ are independently zero, as we have done.

From the components $E_{\theta_{j}\theta_{n}}^{+}$, $E_{\phi_{j}\phi_{n}}^{+}$, and $E_{\phi_{j}\phi_{j}}^{II}$, we obtain that the fields $H_{jn}^{+}, H_{jn}^{\oplus}$ and $H_{j}^{\oplus}$ obey the
P\"oschl-Teller equation, namely
\begin{equation}\label{PoschlTeller2}
  \frac{d^2}{dx^2}H + \left[ \omega^2 - \frac{\Lambda\, (\kappa - 2) }{\cosh^2(\sqrt{\Lambda}\,x )}\right] H = 0 \,.
\end{equation}
In order to conclude this, we have assumed that Einstein's equation $E_{\phi_{j}\phi_{j}}^{I} =0$ holds. As discussed in the previous section, assuming the suitable boundary conditions for the quasinormal modes of the gravitational perturbation the allowed frequencies are given by
\begin{equation}\label{Spectrum2}
  \omega = \sqrt{\Lambda}\left[\, \sqrt{\kappa -9/4 } \,+\, i \,\left(N+1/2\right) \,\right] \,,
\end{equation}
where $N\in\{0,1,2,\cdots \}$, see Eq. (\ref{Spectrum}).

Now, defining the fields $V_{jn}^{I} = V_{[jn]}^{I}$ and $V_{jn}^{II} = V_{[jn]}^{II}$ as
\begin{eqnarray*}\label{HIAII}
V_{jn}^{I} &:=& \frac{1}{f}\left( \frac{d}{dx} H^{+}_{tj} - i \omega  H^{+}_{xj} - \frac{d}{dx} H^{+}_{tn} + i \omega  H^{+}_{xn}\right) \,,\nonumber\\
V_{jn}^{II} &:=& \frac{1}{f}\left( \frac{d}{dx} H^{+}_{xj} - i \omega  H^{+}_{tj} - \frac{d}{dx} H^{+}_{xn} + i \omega  H^{+}_{tn}\right) \,,
\end{eqnarray*}
and assuming that $E_{\phi_{j}\phi_{j}}^{I} =0$, so that the dependence on $E_{\phi_{j}\phi_{j}}^{I}$ can be omitted from the definitions of $E_{t\theta_{j}}^{+}$ and $E_{x\theta_{j}}^{+}$, the following identities hold:
\begin{align*}
  \frac{d^2}{dx^2}V_{jn}^{I}& + \left[ \omega^2 - \frac{\Lambda\, (\kappa - 2) }{\cosh^2(\sqrt{\Lambda}\,x )}\right] V_{jn}^{I} =\\
  &\frac{d}{dx}\left( E_{t\theta_{j}}^{+} - E_{t\theta_{n}}^{+} \right) - i\omega \left( E_{x\theta_{j}}^{+} - E_{x\theta_{n}}^{+} \right)\,,\\
  \frac{d^2}{dx^2}V_{jn}^{II}& + \left[ \omega^2 - \frac{\Lambda\, (\kappa - 2) }{\cosh^2(\sqrt{\Lambda}\,x )}\right] V_{jn}^{II} =\\
  &\frac{d}{dx}\left( E_{x\theta_{j}}^{+} - E_{x\theta_{n}}^{+} \right) - i\omega \left( E_{t\theta_{j}}^{+} - E_{t\theta_{n}}^{+} \right) \,.\\
\end{align*}
Thus, whenever the field equations $E_{t\theta_{j}}^{+} = 0$ and $E_{x\theta_{j}}^{+} = 0$ hold, it follows that $V_{jn}^{I}$ and $V_{jn}^{II}$ obey the P\"oschl-Teller equation and, therefore, have the spectrum (\ref{Spectrum2}).

It is worth recalling that our gauge choice is such that $H_{t2}^{+} = H_{x2}^{+} = 0$. So, from the identities $E_{t\theta_{j}} - E_{t\theta_{2}} = 0$ and $E_{x\theta_{j}} - E_{x\theta_{2}} = 0$, and assuming that $E^{I}_{\phi_j\phi_j} =0$, it follows that
\begin{eqnarray}
H_{tj}^{+} &=&\frac{1}{\Lambda (\kappa - 2)} \left(\frac{d}{dx} V_{2j}^{I} + i\omega V_{2j}^{II}   \right) \,,\nonumber\\
H_{xj}^{+} &=& \frac{1}{\Lambda (\kappa - 2)} \left(\frac{d}{dx} V_{2j}^{II} + i\omega V_{2j}^{I}   \right)\,,
\end{eqnarray}
and thus the spectrum associated to these fields must be the same as that for $V_{jn}^{I}$ and $V_{jn}^{II}$, namely (\ref{Spectrum2}).

It remains to obtain the spectrum of the fields $H_{tt}, H_{tx}$, and $H_{xx}$. From $E_{tx}^{+} = 0$, we can isolate $H_{tx}$ which leads to the equation
\begin{align*}
  H_{tx}\,=&\,\frac{1}{\kappa}\sum_{j}\left[ \kappa_j f \frac{d}{dx}\left( \frac{1}{f}H_{tj}^+ \right) \right. \\
  &\quad\quad  + i\omega \kappa_j H_{xj}^+
  \left.+ 2i\omega \sqrt{f}\frac{d}{dx}\left( \frac{1}{\sqrt{f}}H_j^{\oplus} \right)   \right] \,.
\end{align*}
Thus, $H_{tx}$  can be written in terms of $H_{tj}^+, H_{xj}^+$, $H_{j}^{\oplus}$  and its derivatives, so that $H_{tx}$ must have same spectrum of the latter fields. Finally, from the equation $E^{+}_{t\theta_2} =0$, we find that
$$  H_{tt} + H_{xx} = \frac{2}{i\omega }\left[\frac{d}{dx} H_{tx} - \Lambda f \sum_{n}\left(i\omega H_{n}^{\oplus} + \kappa_{n} H_{tn}^{+}  \right)  \right]\,, $$
where it has been used that $H_{t2}^{+}$ and $H_{x2}^{+}$ both vanish, due to our gauge choice, and it has been assumed that $E^{I}_{\phi_j\phi_j} =0$, as done before. Now, from the equation $E^{I}_{\phi_2\phi_2} =0$ we find
\begin{equation*}
  H_{tt} - H_{xx} = 2 \Lambda f  \sum_{n\geq3} \left[ H_{n}^{\oplus}  + \kappa_{n} (H_{2n}^+  )  \right] \,.
  \end{equation*}
Thus, from these equations for $H_{tt} \pm H_{xx}$ we conclude that these fields are written in terms of fields that we already proved that have the spectrum (\ref{Spectrum2}). This finishes the proof that in the generalized Nariai spacetime all degrees of freedom of the gravitational perturbation,  scalar, vectorial and tensorial, even and odd, have the same spectrum of quasinormal modes. In particular, when $D = 4$ we have $\kappa = \ell(\ell+1)$ and the spectrum can be written as
\begin{equation}
\frac{\omega}{\sqrt{\Lambda}} = \pm \sqrt{(\ell + 2)(\ell - 1) -\frac{1}{4} } \,+\, i \left(N+\frac{1}{2}\right) \,,
\end{equation}
this is exactly the spectrum of frequencies shown by V. Cardoso in Ref. \cite{Cardoso03}, in which an exact expression for the quasinormal modes of gravitational perturbations of a near extremal Scwarzschild-de Sitter black hole, in four dimensions, was displayed. It is well-known that the extremal limit of the Scwarzschild-de Sitter solution, when the black hole horizon coalesces with the cosmological horizon, yield the Nariai spacetime \cite{Nariai1,Nariai2}, so that both spectra should, indeed, coincide. Nevertheless,  when $D = 4$, our analytical spectrum is in disagreement with the quasinormal frequencies for the tensorial degrees of freedom  of the gravitational perturbation displayed in Refs. \cite{Vanzo, Ortega09}. We believe that this difference might have come from a typo in Ref. \cite{Vanzo} that was replicated in Ref. \cite{Ortega09}.

In order to obtain the spectrum of the even part of the gravitational perturbation it was not necessary to use all field equations displayed in Eq. (\ref{Emais}). More precisely, we have not solved $E_{tt}^+=0$, $E_{xx}^-=0$, and $E_{\phi_n\phi_n}^{I}=0$ for $n>2$. Therefore, it is prudential to check if these remaining equations are consistent with the solutions of the ones that we have used. After some algebra, we have checked that this consistency holds indeed. Thus, once we assume that $H_{jn}^{+}, H_{jn}^{\oplus}$, $H_{j}^{\oplus}$, $V_{jn}^{I}$, and $V_{jn}^{II}$ obey the P\"oschl-Teller equation (\ref{PoschlTeller2}), and that $H_{tj}^+$, $H_{xj}^+$, $H_{tx}^+$, $H_{tt}^+$, and $H_{xx}^+$ are given by the expressions given above, it follows that the remaining components of Einstein's equation are automatically satisfied.


\section{Conclusions}\label{Sec.Conclusion}

In this article we have explored the quasinormal modes of a higher-dimensional generalization of the Nariai spacetime, with dimension $2d$, and obtained that all degrees of freedom of the gravitational perturbation have the same spectrum, namely the one displayed in Eq. (\ref{Spectrum2}). This differs, for example, from what happens in other higher-dimensional spacetimes like Schwarzschild and (Anti) de Sitter \cite{Konoplya:2003dd,Cardoso:2003qd,Natario:2004jd}, in which different parts of the gravitational perturbation have different spectra. Thus, the isospectral property of the higher-dimensional the Nariai spacetime considered here proves that the existence of different spectra to different degrees of freedom of the gravitational field is much more related to the symmetries of the spacetime than to the tensorial nature of the degree of freedom of the perturbation or to the dimension of the background. Here the background has $SO(3)\times SO(3)\times \cdots \times SO(3)$ symmetry, $d-1$ times, whereas Schwarzschild black hole has a $SO(2d-1)$ symmetry.

The angular basis constructed here can also be used to separate the degrees of freedom of the gravitational perturbation propagating on other backgrounds with the symmetry $SO(3)\times SO(3)\times \cdots \times SO(3)$. In particular, the higher-dimensional black hole presented in Ref. \cite{Carlos1} can certainly be handled with the technique introduced here. The same idea can also be applied to any spacetime that is the direct product of several spaces of constant curvature.

This work completes the results obtained in a previous paper in which the quasinormal spectrum for fields with spin $0$, $1/2$, and $1$ have been explicitly calculated \cite{JoasI}. Now, with all these results at hand, we can write down a unique formula that works for all of these cases:
$$  \omega = \sqrt{\Lambda}\left[\, \sqrt{\mu^2 + L^2 - \left( s-1/2\right)^2 } \,+\, i \,\left(N+1/2\right) \,\right] \,,$$
where $s$ is the spin of the perturbation, $\mu$ is the mass of the field, and $L^2$ is the squared angular momentum eigenvalue. For instance, the gravitational perturbation amounts to choosing $\mu=0$, $s=2$, and $L^2 = \kappa = \sum_j \ell_j(\ell_j+1)$. The expression for $L^2$ is the same for the scalar field ($s=0$) and for the electromagnetic field ($s=1$), since these are all bosonic fields. On the other hand, for the spin $1/2$ field we have $L^2 =\sum_j \lambda_j^2 $, where $\lambda\in\{\pm1,\,\pm2,\,\cdots\}$ are the eigenvalues of the Dirac operator on the unit sphere. It is worth pointing out that while in Ref. \cite{JoasI} Einstein's vacuum equation was not assumed to hold, so that the spheres of the generalized Nariai spacetime could have different radii \cite{Carlos1}, depending on the electromagnetic charges of the background, here we have assumed vanishing charges, so that the gravitational perturbation decouples from the electromagnetic perturbation. Otherwise, we would have to consider the gravitational and electromagnetic perturbations simultaneously, since the electromagnetic perturbation field would be a source for the gravitational perturbation, as discussed above in Sec. \ref{Sec.Problem}.

\vspace{0.5cm}

\begin{acknowledgments}
C. B. would like to thank Conselho Nacional de Desenvolvimento Cient\'{\i}fico e Tecnol\'ogico (CNPq) for the partial financial support through the research productivity fellowship. Likewise,  C. B. thanks Universidade Federal de Pernambuco for the funding through Qualis A project.  J. V. thanks CNPq for the financial support. We both acknowledge the relevance of CAPES (Coordena\c{c}\~{a}o de Aperfei\c{c}oamento de Pessoal de N\'{\i}vel Superior) for supporting the graduation program at UFPE.
\end{acknowledgments}


\appendix

\section{Basis for the Angular Functions }\label{Sec.Appendix}

In order to facilitate the attainment of the results we provide the explicit expressions for the basis of tensor fields adapted to the spherical symmetry that we have used to span the various degrees freedom of the gravitational perturbation in the generalized Nariai spacetime.

\begin{align*}
  V^{+} &= \hat{\nabla}_{a}Y_{\ell}^{m} \,dx^a = \partial_\theta Y_{\ell}^{m}d\theta +  \partial_\phi Y_{\ell}^{m}d\phi  \\
V^{-} &= \hat{\epsilon}_{ac}\hat{\nabla}^{c}Y_{\ell}^{m} dx^a =
\csc\theta \, \partial_\phi Y_{\ell}^{m} \, d\theta -\sin\theta \, \partial_\theta Y_{\ell}^{m} \, d\phi \\
T^\oplus &= Y_{\ell}^{m}d\theta^2 + \sin^2\theta \,  Y_{\ell}^{m} d\phi^2 \\
T^+ &= \partial_\theta^2 Y_{\ell}^{m}d\theta^2 +
2\left( \partial_\theta \partial_\phi Y_{\ell}^{m}- \cot\theta\, \partial_\phi Y_{\ell}^{m} \right) \, d\theta d\phi  \\
 &\;\;+ \left( \partial_\phi^2 Y_{\ell}^{m} + \cos\theta\,\sin\theta\, \partial_\theta Y_{\ell}^{m} \right) \, d\phi^2 \\
T^- &= 2\csc\theta \left(\partial_\theta  \partial_\phi  Y_{\ell}^{m} - \cot\theta\, \partial_\phi Y_{\ell}^{m} \right)  d\theta^2 \\
&\;\; + 4\left(\csc\theta \, \partial_\phi^2 Y_{\ell}^{m} +  \cos\theta \, \partial_\theta Y_{\ell}^{m} -
 \sin\theta \, \partial_\theta^2  Y_{\ell}^{m} \right) \, d\theta d\phi   \\
 &\;\;  + \,2 \left( \cos\theta\, \partial_\phi Y_{\ell}^{m} -  \sin\theta\, \partial_\theta  \partial_\phi  Y_{\ell}^{m} \right) \, d\phi^2 \,.\\
\end{align*}


\begin{thebibliography}{9}


\bibitem{Vishveshwara70a} C. V. Vishveshwara, \textit{Scattering of Gravitational Radiation by a Schwarzschild Black-hole}, Nature \textbf{227} (1970), 936.

\bibitem{Abbott} B. P. Abbott et al. (LIGO Scientific and Virgo Collaborations), \textit{Observation of gravitational waves from a binary black hole merger}, Phys. Rev. Lett. \textbf{116} (2016), 061102.

\bibitem{Konoplya:2016pmh}
  R.~Konoplya and A.~Zhidenko, \textit{Detection of gravitational waves from black holes: Is there a window for alternative theories?}, Phys.\ Lett.\ B {\bf 756} (2016), 350.

\bibitem{Berti09} E. Berti, V. Cardoso, and A. O. Starinets, \textit{Quasinormal modes of black holes and black branes}, Classical Quantum Gravity \textbf{26} (2009), 163001.

\bibitem{Konoplya:2011qq} R.~A.~Konoplya and A.~Zhidenko, \textit{Quasinormal modes of black holes: From astrophysics to string theory}, Rev.\ Mod.\ Phys.\  {\bf 83} (2011), 793.

\bibitem{Kokkotas} K.D. Kokkotas and B. G. Schmidt, \textit{Quasinormal modes of stars and black holes}, Living Rev. Relativ. \textbf{2} (1999), 2.

\bibitem{Nollert99} H. P. Nollert, \textit{Quasinormal modes: the characteristic 'sound' of black holes and neutron stars}, Class. Quantum Grav. \textbf{16} (1999).

\bibitem{Cardosotese} V. Cardoso, \textit{Quasinormal modes and gravitational radiation in black hole spacetimes},  Doctoral thesis (2004), Universidade T\'ecnica de Lisboa. arXiv:gr-qc/0404093.


\bibitem{NollertNnumeric} H. P. Nollert, \textit{Quasinormal modes of Schwarzschild black holes:  The determination of quasinormal frequencies
with very large imaginary parts}, Phys. Rev. D \textbf{47}, 5253 (1993). 

\bibitem{Natario:2004jd} J.~Natario and R.~Schiappa, \textit{On the classification of asymptotic quasinormal frequencies for d-dimensional black holes and quantum gravity}, Adv.\ Theor.\ Math.\ Phys.\  {\bf 8} (2004) no.6,  1001.









\bibitem{Regge} T. Regge and J. A. Wheeler, \textit{Stability of a Schwarzschild Singularity}, Physical Review D \textbf{108} (1957), 1063.

\bibitem{Manasse} F. K. Manasse, \textit{Distortion in the Metric of a Small Center of Gravitational Attraction due to its Proximity to a Very Large Mass}, Journal of Mathematical Physics \textbf{4} (1963), 746.

\bibitem{Brill} D. R. Brill and J. B. Hartle, \textit{Method of the Self-Consistent Field in General Relativity and its Application to the Gravitational Geon},  Physical  Review \textbf{135} (1964),  B271.

\bibitem{Vishveshwara67} C. V. Vishveshwara, \textit{The Stability of the Schwarzschild Metric}, Doctoral thesis, University of Maryland (1967).

\bibitem{Vishveshwara70b} L. Edelstein and C. V. Vishveshwara, \textit{Differential Equations for Perturbations on the Schwarzschild Metric}, Physical  Review \textbf{D} \textbf{1} (1970), 3514.

\bibitem{Zerilli} F. J. Zerilli, \textit{Effective potential for even-parity Regge-Wheeler gravitational perturbations equations}, Physical Review Letters \textbf{24} (1970), 737.

\bibitem{Fackerell} E. D. Fackerell, \textit{Solutions of Zerilli's equations for even-parity gravitational perturbations}, The Astrophysical Journal \textbf{166} (1971), 197.




\bibitem{Moncrief01} V. Moncrief, \textit{Stability of Reissner-Nordstr\"om}, Physical Review D \textbf{10} (1974), 1057.

\bibitem{Moncrief02} V. Moncrief, \textit{Odd-parity stability of a Reissner-Nordstr\"om black hole}, Physical Review D \textbf{9} (1974), 2707.

\bibitem{Takahashi} T. Takahashi and J. Soda, \textit{Master Equations for Gravitational Perturbations of Static Lovelock Black Holes in Higher Dimensions}, Progress of Theoretical Physics \textbf{124} (2010), 5.

\bibitem{Kodama} H. Kodama and A. Ishibashi, \textit{Master equations for perturbations of generalized static black holes with charge in higher dimensions},   Prog.\ Theor.\ Phys.\  {\bf 111} (2004), 29.

\bibitem{Teukolsky72} S. A. Teukolsky, \textit{Separable Wave Equations for Gravitational and Electromagnetic Perturbations},  Physical Review Letters \textbf{29} (1972), 1114.

\bibitem{Teukolsky74} S. A. Teukolsky, \textit{Separable Wave Equations for Gravitational and Electromagnetic Perturbations}, The Astrophysical Journal \textbf{193} (1974), 443.

\bibitem{Barragan1} J. B. Amado, B. Carneiro da Cunha, and E. Pallante, \textit{On the Kerr-AdS/CFT correspondence}, JHEP 08 (2017), 094.

\bibitem{Barragan-Amado:2018pxh}
  J.~Barragán Amado, B.~Carneiro Da Cunha and E.~Pallante, \textit{Scalar quasinormal modes of Kerr-AdS5}, Phys.\ Rev.\ D {\bf 99} (2019) no.10,  105006.







\bibitem{Carlos1} C. Batista, \textit{ Generalized charged Nariai solutions in arbitrary even dimensions with multiple magnetic charges}, Gen.
Relativ. Gravit. \textbf{48} (2016), 160.


\bibitem{VitorN} V. Cardoso, O. J. C. Dias, J. P. S. Lemos, \textit{Nariai, Bertotti-Robinson and anti-Nariai solutions in higher dimensions}, Phys. Rev. D \textbf{70} (2004), 024002.

\bibitem{Nariai1} P.H. Ginsparg and M. J. Perry, \textit{Semiclassical perdurance of de sitter space}, Nucl. Phys. B 222 (1983), 245.

\bibitem{Nariai2} R. Bousso and S. W. Hawking, \textit{Pair creation of black holes during inﬂation}, Phys. Rev. D 54 (1996), 6312.
arXiv:gr-qc/9606052




\bibitem{Cardoso03} V. Cardoso, \textit{Quasinormal modes of the near extremal Schwarzschild-de Sitter black hole},  Phys. Rev. D \textbf{67} (2003), 084020.



\bibitem{Vanzo} Vanzo, L. and Zerbini, S., \textit{Asymptotics of quasi-normal  modes  for  multi-horizon  black  holes},  Phys.  Rev.  D  \textbf{70} (2004), 044030. arXiv:hep-th/0402103.

\bibitem{Ortega09} A. Lopez Ortega, \textit{The Dirac equation in D-dimensional spherically symmetric spacetimes}, Lat. Am. Jour. Phys. Educ. \textbf{3} (2009), 578. arXiv:0906.2754.

\bibitem{JoasI} J. Ven\^ancio and C. Batista, \textit{Quasinormal Modes in Generalized Nariai Spacetimes}, Physical Review D \textbf{97} (2018), 105025.

\bibitem{Teller} G. P\"oschl and E. Teller, \textit{Bemerkungen zur Quantenmechanik des anharmonis-
chen Oszillators}, Zeitschrift f\"ur Physik \textbf{83} (1933), 143.

\bibitem{Konoplya:2003dd} R.~A.~Konoplya, \textit{Gravitational quasinormal radiation of higher dimensional black holes}, Phys.\ Rev.\ D {\bf 68} (2003), 124017.


\bibitem{Cardoso:2003qd} V.~Cardoso, J.~P.~S.~Lemos and S.~Yoshida, \textit{Scalar gravitational perturbations and quasinormal modes in the five-dimensional Schwarzschild black hole}, JHEP {\bf 0312} (2003), 041.






































\end{thebibliography}
\end{document}